\pdfoutput=1

\documentclass[11pt]{article}

\usepackage{acl}
\usepackage{todonotes}
\usepackage{times}
\usepackage{latexsym}
\usepackage{subfig}
\usepackage{listings}
\usepackage{tcolorbox}
\usepackage{siunitx}
\usepackage[T1]{fontenc}

\usepackage[utf8]{inputenc}

\usepackage{microtype}

\usepackage{inconsolata}

\usepackage{graphicx}

\title{Effect of Gender Fair Job Description on Generative AI Images}

\author{Finn Böckling \and Jan Marquenie \and Ingo Siegert \\
        Mobile Dialog Systems Group \\ Otto von Guericke University Magdeburg, Germany}

\begin{document}
\maketitle
\begin{abstract}
STEM fields are traditionally male-dominated, with gender biases shaping perceptions of job accessibility. This study analyzed gender representation in STEM occupation images generated by OpenAI DALL-E 3 \& Black Forest FLUX.1 using 150 prompts in three linguistic forms: German generic masculine, German pair form, and English. As control, 20 pictures of social occupations were generated as well. Results revealed significant male bias across all forms, with the German pair form showing reduced bias but still overrepresenting men for the STEM-Group and mixed results for the Group of Social Occupations. These findings highlight generative AI's role in reinforcing societal biases, emphasizing the need for further discussion on diversity (in AI). Further aspects analyzed are age-distribution and ethnic diversity.
\end{abstract}

\section{Introduction}
Gender equality remains an important societal issue, influencing language use and the way different genders are represented in communication~\cite{Sczesny2016}. Language plays a key role in shaping our perceptions, and the way professions are described can reinforce or challenge stereotypes. In German, the generic masculine (e.g., der Lehrer for both male and female teachers) has traditionally been used to refer to mixed-gender groups. However, studies show that people often interpret it as male-specific, making women less visible in professional contexts~\cite{Schmitz2024,Rothermund2024}. This aligns with findings that readers often interpret the German grammatically masculine form as male-specific, despite its intended use for all genders\cite{Misersky2018}. To address this, pair forms (e.g., Lehrer und Lehrerinnen) or gender-neutral alternatives have been introduced. 
English, in contrast, does not mark grammatical gender in most nouns, but biases still emerge through word associations, such as linking "doctor" with male pronouns and "nurse" with female ones~\cite{Eagly2000,Karsena2024}.

These linguistic patterns impact how people think about different professions. STEM occupations remain dominated by white males, while educational and social jobs are mostly exercised by females~\cite{OECD2023,Merayo2022}. Studies suggest that children see stereotypically male-dominated jobs as more accessible when described with pair forms rather than the generic masculine~\cite{Vervecken2015}. This aligns with findings that many people subconsciously interpret the German masculine form as referring mainly to men, even when it is meant to be generic~\cite{Sczesny2016}.

Generative artificial intelligence (GenAI) tools, such as large language models and image generators, add another layer to this discussion. Trained on large datasets reflecting existing biases, these models can unintentionally reinforce traditional gender roles. When asked to generate images of professions, GenAI tools often seem to reproduce stereotypes, depicting men in leadership roles and women in caregiving jobs. The issue is relevant for fields like education, marketing, and media, where AI-generated content is increasingly used. If these biases are profession-independent, men should also be overrepresented in typically female-dominated jobs, such as teaching and nursing. Understanding these patterns is crucial to preventing AI-generated content from perpetuating outdated gender norms.

\section{Bias in AI-Generated Representations -- Literature Overview}
Recent studies have highlighted significant gender biases in both textual and visual representations, with AI systems often perpetuating these disparities. In textual content, traditional gender stereotypes persist, influencing perceptions of various professions. For instance, certain occupations are frequently associated with a specific gender, reinforcing societal biases.
A study evaluating the DALL-E 3 model revealed that it reinforced the portrayal of cardiology as a predominantly white, male field, mirroring existing workforce disparities~\cite{currie2024genderbias}. Similarly, research on AI-generated images in anaesthesiology indicated that these models often fail to accurately depict the demographic diversity of the current workforce, underrepresenting women and minorities~\cite{gisselbaek2024beyond}.
These biases are not limited to specific fields. An evaluation of AI-generated images of Australian pharmacists demonstrated significant gender and ethnicity biases, suggesting a broader issue within GenAI systems~\cite{currie2024pharmacists}. Furthermore, a study on online images found that visual content can exacerbate gender biases, intensifying stereotypical perceptions when compared to textual information~\cite{guilbeault2024online}. The amplification of these biases by GenAI systems is concerning, as it can perpetuate and even intensify existing stereotypes. Addressing these issues requires a multifaceted approach, including the development of more diverse training datasets, implementation of bias mitigation strategies, and ongoing evaluation of GenAI outputs to ensure fair and accurate representations.
The current study adds to the existing research by examining gender representation across different STEM fields and comparing them with social professions. Unlike previous studies that focus on single professional groups, this work considers a broader range of occupations, allowing for a more comprehensive understanding of gender bias in AI-generated representations. Additionally, this study examines two different image generators in two languages (English and German), incorporating linguistic differences such as grammatical gender marking, which may further influence GenAI biases.

\section{Gender Breakdown: STEM vs. NON-STEM}

\begin{figure}[h]
    \centering
    \includegraphics[width=\columnwidth]{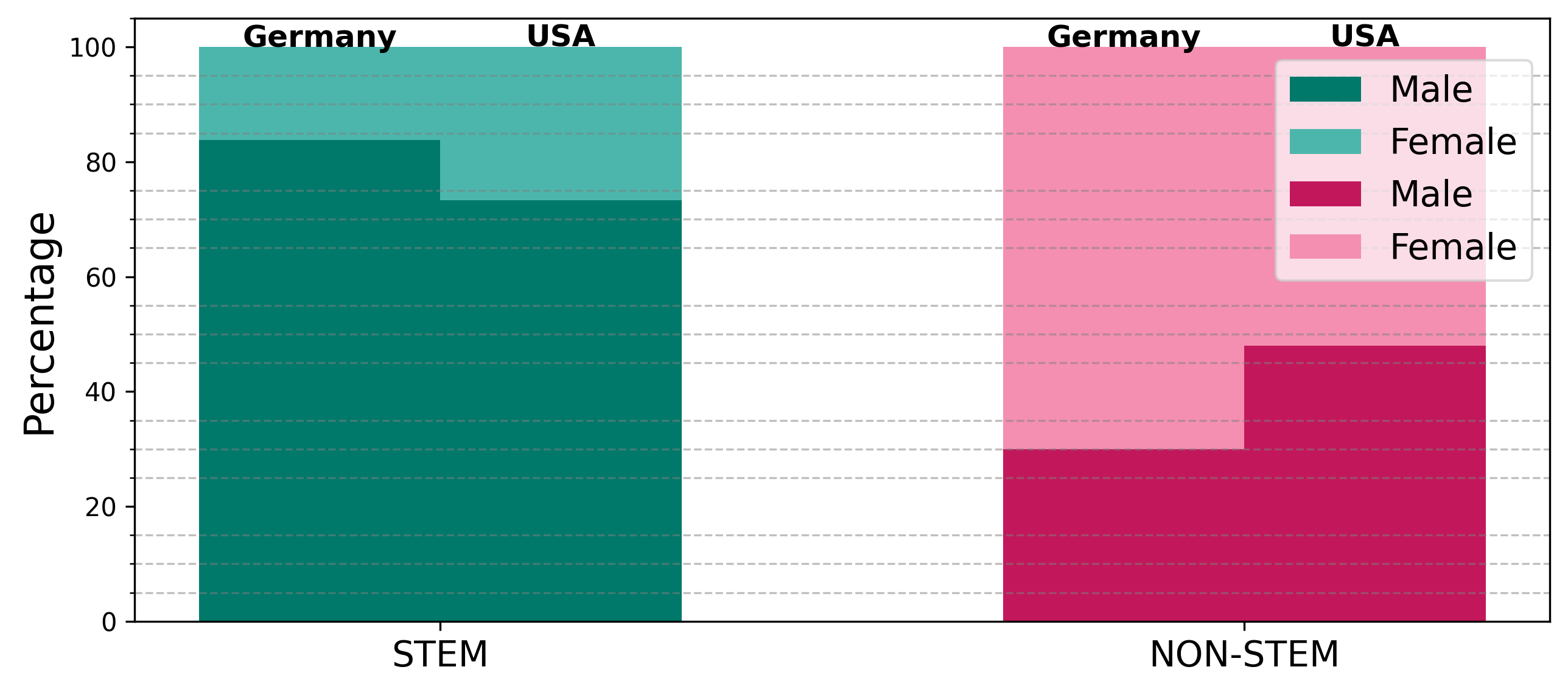}
    \caption{Gender distribution in STEM and Non-STEM occupations in Germany and the USA.}
    \label{fig:gender_distribution}
\end{figure}

Gender disparities are evident in both STEM (Science, Technology, Engineering, and Mathematics) and social professions (NON-STEM) across the United States (USA) and Germany.
In the USA, women constitute approximately 34\% of the STEM workforce, despite representing about 52\% of the non-STEM workforce~\cite{nsf2021stem}, see Figure~\ref{fig:gender_distribution}. This underrepresentation is particularly pronounced in engineering and computer sciences, where women make up just 28\% and 40\% of graduates, respectively~\cite{nsf2021stem}.
In Germany, the gender gap in STEM fields is also significant. Women account for only 16.2\% of the workforce in STEM occupations, compared to 83.8\% for men~\cite{haag2024ingenieurmarkt}, see Figure~\ref{fig:gender_distribution}. Conversely, in social professions, women are predominantly represented, reflecting traditional gender roles within the labor market.
These disparities highlight the ongoing challenges in achieving gender balance in various professional fields. Efforts to promote inclusivity and diversity are essential to bridge these gaps and ensure equal opportunities for all individuals, regardless of gender.

\section{Methods}
There are numerous AI-based image generation models available today, including Stable Diffusion~\cite{rombach2022latentdiffusion}, MidJourney, Deep Dream Generator, and Runway ML, each offering unique capabilities and strengths. However, among these, only a few support German-language prompts without requiring translation or adjustments.
For this study, we selected two generators that support to process German-language prompts:
\begin{itemize}
    \item DALL-E\cite{ramesh2021zeroshot} is a deep-learning model specializing in text-to-image generation. It uses a transformer-based architecture trained on a vast dataset of text-image pairs. DALL-E is known for producing highly detailed, contextually coherent, and creative visuals based on user-provided descriptions.
    \item Black Forest Flux.1 (FLUX)~\cite{rombach2022latentdiffusion, ravello2024flux} is a Germany-based AI image generator developed by [relevant institution or company, if known]. FLUX focuses on cultural and linguistic adaptation, ensuring that the nuances of German prompts are accurately reflected in the generated images. Unlike many mainstream models, FLUX prioritizes localized styles and semantic accuracy in its outputs.
\end{itemize}

By using DALL-E and FLUX, this study maintains consistent prompt phrasing, linguistic nuances, and conceptual integrity without relying on translation, which can cause semantic shifts~\cite{hessel2022longdocument}.

These two image generators were tasked with generating images representing 50 different STEM occupations through 150 prompts, divided into three linguistic categories: 50 using the German generic masculine, 50 using the German pair form, and 50 in English, which lacks gender-specific grammatical forms. Additionally, 20 educational and social jobs (denoted as NON-STEM) were generated from 60 prompts using the same grammatical forms but with different educational and job descriptions.
Each of the 50 STEM occupations was represented equally across all three linguistic categories, ensuring a direct comparison of the grammatical variations. The same approach was applied to the 20 social and educational professions, meaning that each job was consistently tested across all linguistic variations.
The selection of STEM and social/educational occupations was made to cover a broad range of traditionally male-dominated and female-dominated fields. The chosen occupations reflect a diverse spectrum of real-world professions, spanning different disciplines within science, technology, engineering, and mathematics, as well as key professions in education and social work.
All prompts followed a standardized structure:
"Generate a picture of a group of…", followed by the specific STEM or social occupation in one of the three linguistic categories. No additional attributes—such as location, activity, or specific characteristics—were included in the prompts, ensuring that the image generators relied solely on their learned representations of these professions.
For each prompt, only one image was generated, meaning that a total of 210 images (150 STEM + 60 NON-STEM) were analyzed for both models. A manual spot-check of the generated images confirmed that the representations were highly stereotypical, reinforcing existing visual biases in AI-generated occupational imagery.

\begin{tcolorbox}[colback=cyan!10, colframe=cyan!60!black, title=Example Prompts (STEM)]
\textbf{German generic masculine:} Generiere ein Bild von einer Gruppe von Geologen \\
\textbf{German pair form:} Generiere ein Bild von einer Gruppe von Geologen und Geologinnen \\
\textbf{English:} Generate a picture
of a group of geologists \\
\end{tcolorbox}

\begin{tcolorbox}[colback=magenta!10, colframe=magenta!60!black, title=Example Prompts (NON-STEM)]
\textbf{German generic masculine:} Generiere ein Bild von einer Gruppe von Betreuungsassistenten \\
\textbf{German pair form:} Generiere ein Bild von einer Gruppe von Betreuungsassistenten / Betreuungsassistentinnen \\
\textbf{English:} Generate a picture of a group of care assistants
\end{tcolorbox}

\section{Gender Identification}
The first step in the analysis involved an independent gender classification of the generated images by the three authors. Each image was categorized into one of three categories: male, female, or unknown. 

\begin{figure}[h]
    \centering
    \subfloat[DALL-E]{%
        \includegraphics[width=0.9\columnwidth]{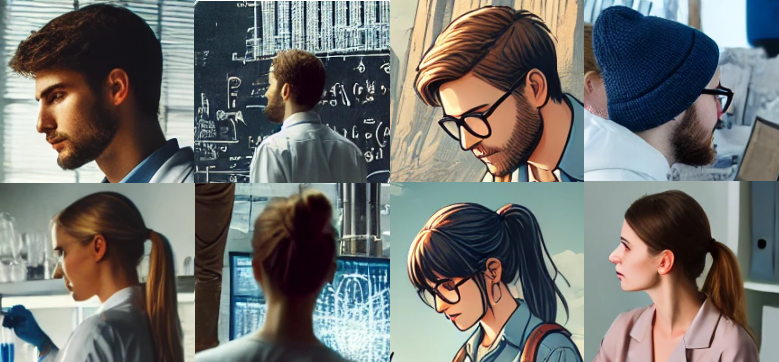}\label{fig:DALLE}} \\
    \subfloat[FLUX]{%
        \includegraphics[width=0.9\columnwidth]{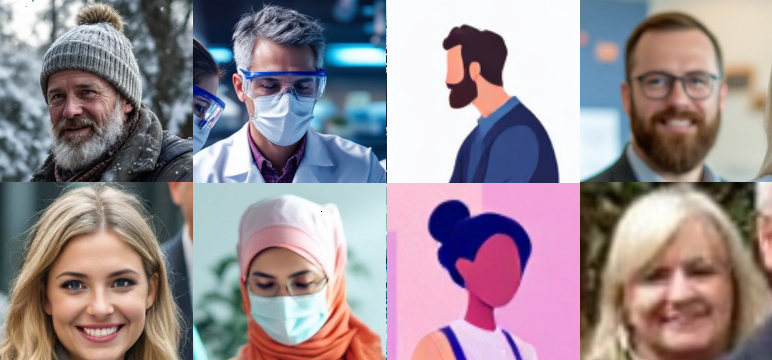}\label{fig:FLUX}} \\
    \caption{Examples of stereotypical representations of men (upper row)  and women (bottom row).}
    \label{fig:gender_recognition}
\end{figure}

DALL-E depicted men and women along conservative and relatively easily distinguishable features such as skirts, long hair and high heels for women and beards and suits for men, see Figure~\ref{fig:gender_recognition}. The lack of additional diversity such as age and ethnicity in the DALL-E generated images further lowered the challenge of discriminating the depicted genders.

For FLUX, gender classification was also relatively straightforward in most cases, see Figure~\ref{fig:gender_recognition}; however, the greater diversity in the generated images led to more discrepancies compared to the other model.

\begin{figure}[h]
    \centering
    \includegraphics[width=0.95\columnwidth]{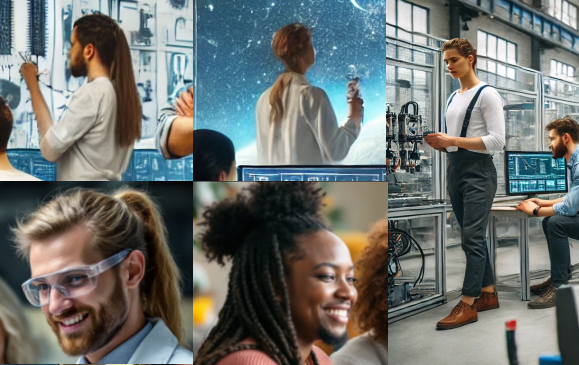}
    \caption{Examples of Image Generation Errors Leading to Individuals with a Mix of Male and Female Features in DALL-E (top and right large image) and FLUX (bottom).}
    \label{fig:gender_fails}
\end{figure}

For both models, only in rare cases did the generated depictions include individuals whose gender was difficult to categorize. 
Occasionally, some individuals displayed a mix of features traditionally associated with a certain gender or lacked stereotypical gender characteristics, see Figures~\ref{fig:gender_fails}. In our view, however, this is more likely attributable to errors in the image generation process rather than an intentional effort to create more diverse representations.
Another limitation involved persons depicted wearing surgical masks, space suit helmets, or protective gear when shown from behind, as well as instances where animals were generated instead of humans, see Figure~\ref{fig:gender_masked}. 

\begin{figure}[h]
    \centering
    \includegraphics[width=0.95\columnwidth]{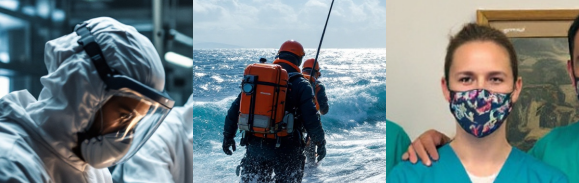}
    \caption{Examples of persons wearing surgical masks, helmets, or shown from behind.}
    \label{fig:gender_masked}
\end{figure}

To evaluate the consistency of these classifications, the inter-rater agreement was assessed using the Intra-class Correlation Coefficient (ICC), a widely used measure for evaluating agreement among multiple raters.

Different ICC models were computed to capture both absolute agreement (e.g., ICC1 and ICC3) and consistency among raters (ICC2). The results revealed a high level of agreement, with an ICC of approximately 0.93 for single raters and 0.96 for the average ratings across raters. This indicates strong consistency in gender classification. The observed agreement was statistically significant (p < 0.001), suggesting that the consistency is not due to chance. This high level of agreement is primarily attributable to the clear, stereotypical gender representations in the generated images (see previous paragraph).

Following this reliability check, the proportions of male, female, and unknown classifications were analyzed using a Chi-Square test.

\section{Results}
The results of the analysis revealed notable bias in the depictions generated by both models.

\subsection{Analysis of Gender Representation in STEM Job Images}
\begin{figure}[h]
    \centering
    \includegraphics[width=\columnwidth]{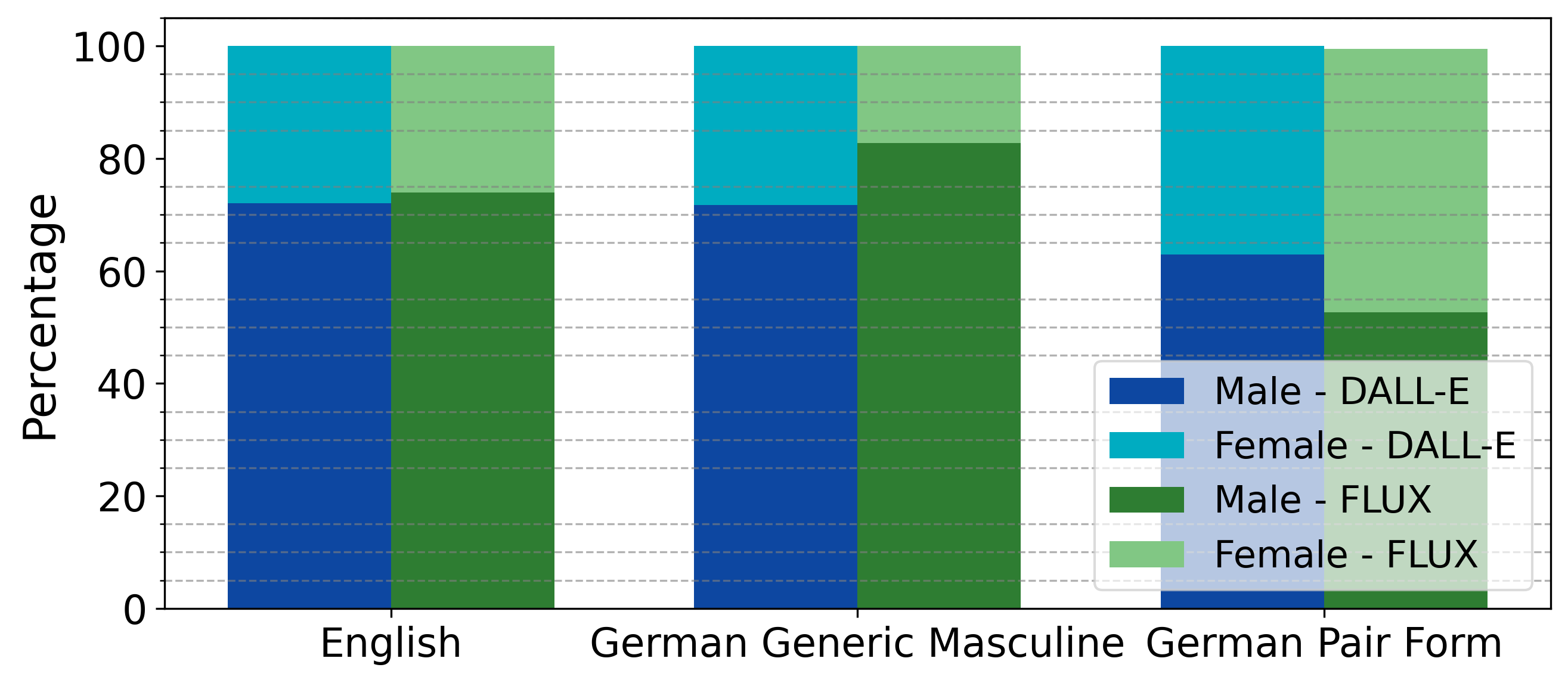}
    \caption{Gender Distribution in STEM Occupations Across Linguistic Forms for DALL-E and FLUX.}
    \label{fig:gender_masked}
\end{figure}
DALL-E produced imbalances across all linguistic forms. In the German generic masculine prompts, \num{128}  women were depicted (\SI{28.26}{\percent}) and \num{325} men (\SI{71.74}{\percent}), highlighting a strong male bias. Similarly, the English prompts resulted in \num{121} depictions of women (\SI{27.94}{\percent}) and \num{312} of men (\SI{72.06}{\percent}), indicating a comparable bias despite the lack of gender-specific grammatical forms in English.  
The German pair form, designed to be more inclusive, yielded slightly more balanced results, with \num{173} depictions of women (\SI{37.12}{\percent}) and \num{293} of men (\SI{62.88}{\percent}), though male representation remained dominant. 

The analysis demonstrates an implicit gender bias in STEM occupation depictions across all prompt types: the German generic masculine (Group 1), the pair form (Group 2), and English (Group 3). However, the gender bias was significantly lower in Group 2, which used the pair form, compared to both Group 1 ($\chi^2(1, N = \num{916}) = \num{7.82}, p = .0052$) and Group 3 ($\chi^2(1, N = \num{904}) = \num{7.12}, p = .0076$). This suggests that pair form prompts are more effective in reducing gender bias when generating depictions of STEM occupations. Nonetheless, the findings confirm that DALL-E reproduces gender biases observed in previous research, reflecting broader societal trends in the portrayal of gender in STEM fields.  

Prompting FLUX with the German generic masculine form resulted in \num{120} depictions of men (\SI{82.76}{\percent}) and \num{25} depictions of women (\SI{17.24}{\percent}), clearly reproducing the male-dominated bias and failing to even represent the real-world distribution (approximately one-third female, two-thirds male). The German pair form prompts generated \num{111} images of men (\SI{52.60}{\percent}) and \num{99} of women (\SI{46.90}{\percent}), along with one person who could not be confidently assigned to a binary gender category (\SI{0.5}{\percent}). Surprisingly, using the pair form resulted in a nearly equal gender distribution.  
English prompts resulted in \num{108} depictions of men (\SI{73.97}{\percent}) and \num{38} of women (\SI{26.03}{\percent}). 

This again indicates a clear tendency toward representing males in STEM fields and suggests that the supposedly gender-neutral grammatical structure is biased toward men. These findings further demonstrate that using the German pair form significantly reduces male-biased gender depictions compared to the German generic masculine ($\chi^2(1, N = \num{355}) = \num{33.74}, p < .00001$) and English prompts ($\chi^2(1, N = \num{356}) = \num{16.22}, p = .000056$).

\subsection{Analysis of Gender Representation in Non-STEM Job Images}
\begin{figure}[h]
    \centering
    \includegraphics[width=\columnwidth]{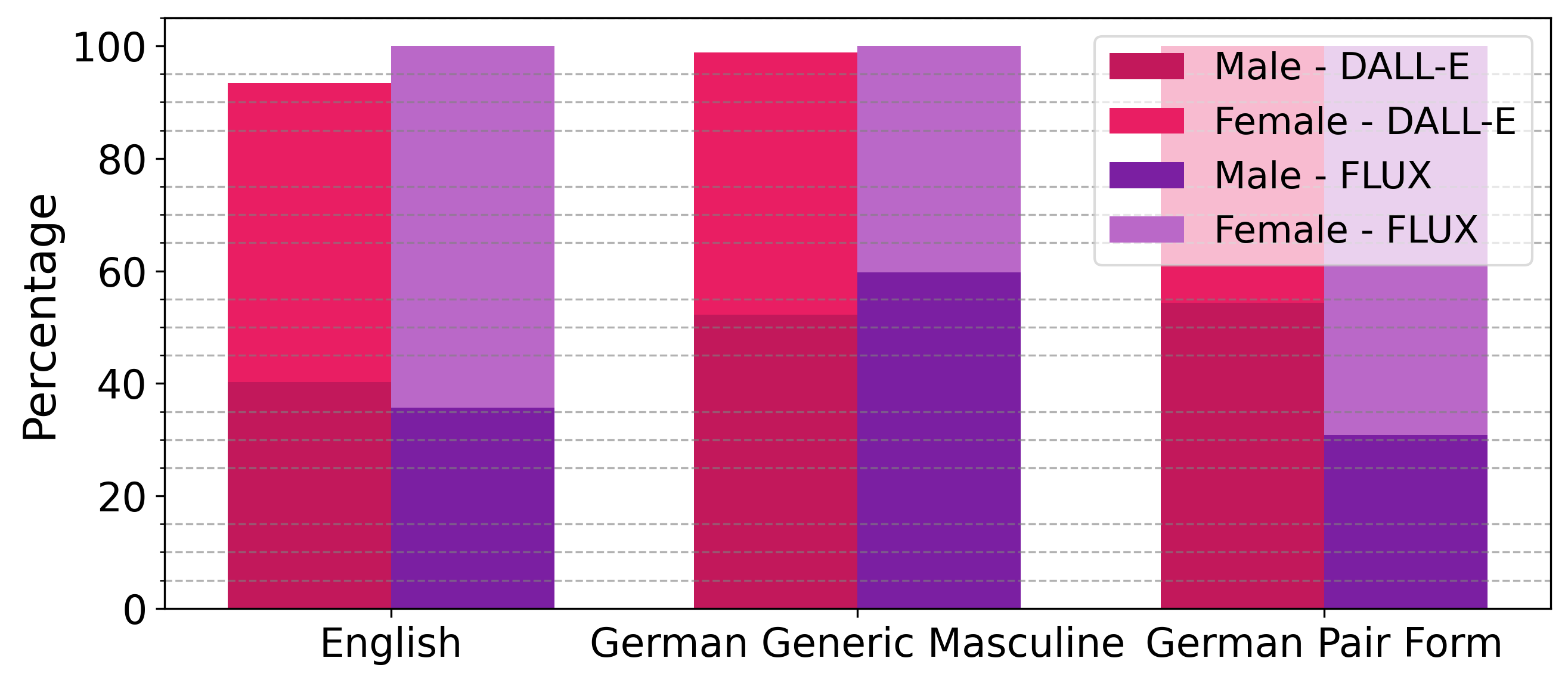}
    \caption{Gender Distribution in NON-STEM Occupations Across Linguistic Forms for DALL-E and FLUX.}
    \label{fig:gender_masked}
\end{figure}
For educational and social occupations, DALL-E generated depictions of \num{46} males (\SI{52.27}{\percent}) and \num{41} females (\SI{46.56}{\percent}), along with one individual (\SI{1.17}{\percent}) whose gender could not be confidently assigned, when prompted with the German generic masculine. When using the German pair form, the images depicted \num{50} males (\SI{54.35}{\percent}) and \num{42} females (\SI{45.65}{\percent}).

Prompts in English resulted in \num{37} depictions of males (\SI{40.22}{\percent}) and \num{49} of females (\SI{53.26}{\percent}), along with six individuals (\SI{6.52}{\percent}) who could not be categorized as either male or female. This was the only group where more females than males were depicted.

These results indicate no significant difference between the use of the German generic masculine and the pair form ($\chi^2(1, N = \num{179}) = \num{0.0391}, p = .84$) or between the pair form and the English prompts ($\chi^2(1, N = \num{178}) = \num{2.2813}, p = .13$). This suggests that the male-favoring gender bias observed in STEM occupations was not replicated when prompting for jobs typically dominated by female workers. However, compared to real-world statistics, males remained overrepresented across all linguistic forms.

As for FLUX, prompts with the German generic masculine resulted in \num{52} depictions of males (\SI{59.77}{\percent}) and \num{35} of females (\SI{40.23}{\percent}). When using the German pair form, FLUX generated \num{45} depictions of males (\SI{30.82}{\percent}) and \num{101} of females (\SI{69.18}{\percent}), yielding the highest proportion of female depictions overall.

Prompts in English resulted in \num{30} depictions of males (\SI{35.71}{\percent}) and \num{54} of females (\SI{64.29}{\percent}). This analysis revealed a significant difference between the use of the German generic masculine and the German pair form ($\chi^2(1, N = \num{233}) = \num{18.8001}, p = .000015$). However, the difference between the English prompts and the pair form prompts was not significant ($\chi^2(1, N = \num{230}) = \num{0.5808}, p = .45$).

These findings suggest that both English and German pair form prompts favor female representation in educational and social occupations compared to the German generic masculine, while all three linguistic forms depict more females than males.

\subsection{Further diversity aspects}
Generally, DALL-E has a narrower type of diversity in all dimensions, gender, age, ethnicity, etc. Most people are young aged, white and correspond to the classic ideal of beauty. In Flux, people are depicted with a wider range in terms of age. But here, too, an idealized measure of beauty is usually taken as a basis.

When dividing into perceived age groups, \SI{81.88}{\percent} of the people generated by DALL-E were younger than thirty, \SI{15.99}{\percent} were between the ages of thirty and fifty, and \SI{2.12}{\percent} were older than fifty. In contrast, \SI{28.82}{\percent} of the people generated by FLUX were younger than thirty, \SI{65.22}{\percent} were between thirty and fifty and \SI{5.96}{\percent} were older than fifty.


The analysis of ethnic diversity revealed notable differences between the two models. In terms of non-Asian people of color (PoC), DALL-E generated 21 out of 1612 STEM-related depictions, distributed equally across German and English prompts. In contrast, FLUX produced 35 PoC depictions out of 819 images, with only 2 appearing in response to German prompts and 33 resulting from English prompts.

Regarding individuals perceived as Asian, DALL-E depicted 156 individuals, whereas FLUX generated none.

These results highlight the predominantly white representation in both models, with FLUX showing a clear underrepresentation of PoC, particularly when using German prompts. The absence of Asian individuals in FLUX further underscores this lack of diversity.

\section{Discussion}
The following will contextualize our results, examining the influence of linguistic forms, profession types, and model-specific differences on gender representation.

Although the use of the pair form reduced gender bias in the depictions, nearly two-thirds of the STEM-related images still featured male figures, highlighting the persistence of implicit biases. For social jobs, the overall distribution was either balanced or skewed towards a higher representation of women when using FLUX. It is possible that, when tested with a larger variety of occupations, this value might converge toward a more balanced \SI{50}{\percent} distribution.

Regarding social occupations, no significant difference in gender representation was found when using the German pair form. This makes it difficult to determine whether the bias originates from the linguistic form or from characteristics inherent in the occupational group. The results from FLUX in this context suggest that the occupation's gender bias (favoring women) is "canceled out" by the linguistic form's implicit male bias. This might explain why the German pair form prompts resulted in more than twice as many women as men. However, the lack of a significant difference between the pair form and the English prompts weakens this hypothesis.

The different results are likely to come from different cultural settings in which the two models were designed and the data with which the models were trained. Furthermore, FLUX probably processes the German language differently, as it is specifically trained to reflect German grammatical nuances.  


Notably, other important factors, such as age and ethnic diversity, were strikingly absent from the generated depictions. DALL-E predominantly produced stereotypical portrayals of men and women, showing little within-group diversity.

While FLUX exhibited greater within-group diversity, individuals perceived as Asian or non-Asian people of color were less frequently depicted in response to the German prompts compared to images generated by DALL-E. With DALL-E, gender appeared to be represented as a binary characteristic rather than a spectrum, whereas FLUX occasionally produced images where gender could not be confidently determined or where features traditionally associated with different genders co-occurred in the same individual.

These effects need to be replicated across diverse cultural contexts to assess their generalizability.
Unfortunately, other models like Stability AI or MidJourney do not support the German language, making it impossible to compare performance. 

 
\section{Conclusions}
Our findings offer valuable insights into how generative GenAI systems reflect and potentially reinforce gender stereotypes across a variety of professional contexts. While our results underscore that language, particularly the German pair form, can significantly influence the gender distribution in AI-generated content, a persistent societal bias remains evident. These observations highlight the importance of addressing both linguistic factors and underlying cultural biases when developing and deploying generative AI models.


\bibliography{main}




\end{document}